\documentstyle[11pt,a4]{article}
\begin{document}

\newcommand{\dd}[1]{\frac{\delta}{\delta #1}}
\mathchardef\Dd="3244
\mathchardef\Ww="3257
\mathchardef\Pp="3250

\begin{titlepage}
\begin{flushright} Preprint KUL-TF-93/26\\
                   June 1993\\
\end{flushright}
\vfill
\begin{center}
{\large\bf Factoring out Free Fields}\\
\vskip 27.mm
{\bf Alex Deckmyn$^1$ and Kris Thielemans}\\
\vskip 1cm
Instituut voor Theoretische Fysica
        \\Katholieke Universiteit Leuven
        \\Celestijnenlaan 200D
        \\B-3001 Leuven, Belgium\\[0.3cm]
\end{center}
\vfill
\begin{center}
{\bf Abstract}
\end{center}
\begin{quote}
\small
For a generic $\Ww$ algebra, we give an algorithmic procedure for factoring
out all fields of dimension $1/2$, both bosonic and fermionic, and some
fields of dimension $1$. This generalizes and makes more explicit the
Goddard-Schwimmer theorem for free fermions.  We also show how the
induced gravity theory for the original $\Ww$ algebra containing the free
fields relates to the theory where the fields are factored out.

\begin{flushright}
                   hepth@xxx/9306129\\
\end{flushright}
\vspace{2mm}
\vfill
\hrule width 3.cm  {\footnotesize
\noindent $^1$ Aspirant N.F.W.O., Belgium.\\}
\normalsize
\end{quote}
\end{titlepage}
\newpage

\section{Introduction}
Some years ago, Goddard and Schwimmer \cite{GS} proved that every
meromorphic conformal theory can be factorized into free fermions (of
spin $1/2$) and a part containing no free fermions.  A consequence
of this is that in the classification of  $\Ww$ algebras, spin $1/2$
fermions need never be considered.  This is very fortunate, since the main
method of constructing a large number of  $\Ww$ algebras, hamiltonian
reduction (see e.g.  \cite{FORTW} for a recent account of the classical
case, and \cite{BO,TdB,ST} for the quantum case), does not generally yield
spin $1/2$ fields.  (Supersymmetric reduction, see \cite{FRS}, does
give weight $1/2$ fields.) However, \cite{GS} does not treat
bosonic fields of weight $1/2$ (symplectic bosons).  Mostly,
it is assumed that these fields, too, can always be factorized, but up to
now this was not yet explicitly proven.  In this letter, we present an {\em
algorithmic} procedure for factoring out of both fermionic and bosonic
fields of weight $1/2$.

It was already noticed in \cite{GS} that in some cases (e.g. the $N=4$
superconformal algebra) spin $1$ bosons can also be decoupled
from a conformal theory. This is certainly not a general property.
The factorization-algorithm presented here gives a criterion to
decide when free bosons can be decoupled.

In the second part of this letter, we extend the results of \cite{STT2}.
There, it was shown how factoring out the fermions from the $N=3,4$
superconformal algebras, links linear and non-linear $N=3,4$ induced
supergravity.  We will show here how all the factorizable fields can in
general be integrated out. Also, the criterion for factorizable $U(1)$
fields is rederived from the Ward identities.

\section{Algorithms for factorisation}
In the following subsections, we will show how various free fields can be
decoupled by adding extra composite terms. This will be done recursively
by removing the highest order pole in the OPE of the free field with the
other fields in the algebra. For the energy-momentum tensor,
the extra terms amount exactly to substracting the usual e.m tensor of
the free field, so the new e.m. tensor is again a good Virasoro tensor (with
a central charge $c$ shifted by minus the central charge of the free
fields).  The other fields of the theory may become non-primary after
redefinition, but it should always be possible to find a new primary basis.
We will not go further into this.

We will use the following convention for the OPE of two fields:
\begin{equation}
A(z)B(w) = \sum_{i\leq h(A,B)}^{} \left[ AB\right] _i(w) \left( z-w\right)
^{-i},
\end{equation}
where $h(A,B)$ is usually the sum of the conformal weights of the fields.
We define the modes $\hat{A}_m$ by\footnote{This is actually a shift in the
index $m$ with respect to the usual definition.  This will of course be
reflected in the commutators.}

\begin{equation}
\hat{A}_m B\equiv \left[ AB\right] _m. \label{modes}
\end{equation}

\subsection{Free fermions}
For completeness, we first rederive the result of \cite{GS} in our
formalism and give an explicit algorithm for the decoupling.
Consider a theory containing a free fermion $\psi$ (we will only
consider the Neveu-Schwarz sector):
\begin{equation}
\left[ \psi \psi \right] _1 = \lambda \,\,.
\end{equation}
We have the commutation relation
\begin{equation}
\hat{\psi }_m \hat{\psi }_n = -\hat{\psi }_n \hat{\psi }_m + \lambda \delta
_{m+n-1}.
\label{com1}
\end{equation}
Now suppose we have already {\em partly} decoupled $\psi $ from a
certain field $A$, more exactly
\begin{equation}
\hat{\psi }_k A=0 \; , \; \forall k\geq n+1.
\label{suppose1}
\end{equation}
We can then add a composite field to $A$:
\begin{equation}
\Delta _n A\equiv -\frac{1}{\lambda }\hat{\psi }_{1-n}\hat{\psi }_n
A, \label{delta1}
\end{equation}
such that condition (\ref{suppose1}) is valid for all $k\geq n$:
\footnote{Note that the regular part of an OPE can be
found from the general relation \cite{bais}
\begin{equation}
\left[ AB\right] _{-n} = \frac{1}{n!}\left[ A^{(n)}B\right] _0.
\end{equation}  }
\begin{equation}
\hat{\psi}_k\left(1 + \Delta _n\right)A = 0 \; , \; \forall k\geq n.
\end{equation}
So $\Pp_n\equiv (1+\Delta _n)$ is the projection operator on the kernel of
$\hat{\psi }_n$.
The procedure (\ref{delta1}) can clearly be iterated, and at the end all
the redefined fields will have a regular OPE with $\psi $, which is the
desired result. The complete projection operator becomes
\begin{equation}
\Pp\equiv \prod_{n} \Pp_n.
\end{equation}
For example, it is very easy to check that the e.m.
tensor gets the expected correction
\begin{equation}
\Delta _2 T =-\frac{1}{2\lambda }\psi '\psi .
\end{equation}
Finally, note that through use of Jacobi identities, one sees that the
redefined fields form a closed algebra, where all fields commute with
$\psi$.

\subsection{Symplectic bosons}
Suppose we have a couple of symplectic bosons
\begin{eqnarray}
\left[ \xi ^+ \xi ^-\right] _1 = \lambda = -\left[ \xi ^- \xi ^+\right]_1
,\nonumber\\
\left[ \xi ^+ \xi ^+\right] _1 = \left[ \xi ^- \xi ^-\right]_1 =0.
\end{eqnarray}
The operators $\hat{\xi}^{\pm}_m$, as in (\ref{modes}),
have commutation relations
\begin{eqnarray}
\hat{\xi} ^{\pm}_m\hat{\xi} ^{\pm}_n &=& \hat{\xi} ^{\pm}_n\hat{\xi}
^{\pm}_m,\nonumber\\
\hat{\xi} ^+_m \hat{\xi} ^-_n&=& \hat{\xi} ^-_n\hat{\xi} ^+_m + \lambda
\delta _{m+n-1}.
\label{com2}
\end{eqnarray}
We can define the operators
\begin{equation}
\Delta ^{\pm}_n
\equiv \sum_{i\geq 1}^{} \frac{(\mp1)^{i}}{i!\lambda ^i}\left(
\hat{\xi} ^{\mp}_{1-n}\right) ^i \left( \hat{\xi} ^{\pm}_n\right) ^i,
\label{delta2}
\end{equation}
or
\begin{equation}
\Pp_n^{\pm}\equiv 1+\Delta ^\pm_n\equiv\; :\exp\left[ \mp\frac{1}{\lambda
} \hat{\xi}^{\mp}_{1-n}\hat{\xi}^{\pm}_{n}\right] :\;\;.
\end{equation}
On any field $A$, the sum (\ref{delta2}) will stop after a finite number of
terms, since for every $n\geq 1$ the conformal weight
of $\hat{\xi }^{\pm}_nA$ is strictly smaller than that of $A$ itself, and
there is a lower bound on the dimension of the fields in the algebra.
If
\begin{equation}
\hat{\xi}^{\pm}_k A = 0,\,\, \forall k\geq n+1, \label{suppose2}
\end{equation}
these operators satisfy the commutation relations
\begin{eqnarray}
\Delta^{\pm}_n\Delta^{\mp}_nA &=& \Delta^{\mp}_n\Delta^{\pm}_nA,\nonumber\\
\hat{\xi }^{\pm}_n \Delta ^{\pm}_nA &=& -\hat{\xi }^{\pm}_nA,\nonumber\\
\hat{\xi }^{\pm}_n\Delta^{\mp}_nA  &=& \Delta ^{\mp}_n\hat{\xi }^{\pm}_nA.
\end{eqnarray}
So the complete projection operator is
\begin{equation}
\Pp_n\equiv \Pp_n^+\Pp_n^-.
\end{equation}

Again, this procedure can be iterated up to $n = 1$, proving that the
symplectic bosons can all be decoupled. Note that since the algorithm is
only based on the commutation relations (\ref{com2}) it will work just as
well for a generic $(\beta ,\gamma )$ system. For a fermionic $(b,c)$
system the result is even simpler, since then the sum (\ref{delta2})
reduces to its first term, with an additional minus sign:
\begin{equation}
\Delta _n^\pm\equiv -\frac{1}{\lambda }\xi ^\mp_{1-n}\xi ^\pm_n.
\end{equation}

\subsection{$U(1)$ currents}
When trying to decouple a $U(1)$ field, we will see that this is not always
possible. In fact, there will be no way to set the first order pole to zero
by adding correction terms. We now have
\begin{equation}
\left[ JJ\right] _2 = \lambda
\end{equation}
and
\begin{equation}
\hat{J}_m\hat{J}_n = \hat{J}_n\hat{J}_m+\lambda (m-1)\delta _{m+n-2}.
\end{equation}
Suppose $\hat{J}_kA=0$, $\forall k\geq n+1$, then it is easy to check that
\begin{equation}
\Pp _n \;\equiv\; : \exp\left[ \frac{1}{(1-n)\lambda }
\hat{J}_{2-n} \hat{J}_n\right] :
\label{delta3}
\end{equation}
is the projection operator on the kernel of $\hat{J}_n$. However, the whole
scheme
breaks down at $n=1$, since there is a factor $(1-n)^{-1}$ that diverges.
Also, it is important to notice that the conformal dimension of
$\hat{J}_1A$ is equal to that of $A$ itself, so $(\hat{J}_1)^iA$ need not
be zero even for very large $i$. From this we
conclude that a sufficient condition for the decoupling of a $U(1)$ current
is
\begin{equation}
\hat{J}_1A = \left[JA\right]_1 =0
\label{condition}
\end{equation}
for {\em all} fields $A$ of the theory. In fact, it is quite easy to show
that (\ref{condition}) is also a necessary condition. For the fields of
conformal dimension $1$ this is obvious. Now consider the lowest dimension
$d$ where (\ref{condition}) is not satisfied. From Jacobi identities
it follows that
\begin{equation}
[J[AB]_n]_1 = [[JA]_1B]_n + [A[JB]_1]_n \label{projection}.
\end{equation}
Thus $\hat{J}_1$ working on any
{\em composite} field of this dimension vanishes. So correction terms can
only be non-composite. Clearly, if $\hat{J}_1$ does not vanish on all $N_d$
(non-composite) primary fields of dimension $d$, we can not make $N_d$
independent linear combinations on which it does. Note that
(\ref{condition}) simply tells us that all fields should have zero
$U(1)$ {\em charge} with respect to $J$.

Of course, we may try to decouple a larger
part of the  $\Ww$ algebra, e.g. a Ka\v{c}-Moody algebra, containing $J$. In
this case, condition (\ref{condition}) need not hold.
In this letter, however, we will restrict ourselves to the case
of $U(1)$ fields.

Finally, note that due to \cite{bais}
\begin{equation}
\left[ AJ\right] _1=\left[ JA\right] _1+\sum_{i\geq 2}
\frac{(-1)^i}{(i-1)!}\partial ^{(i-1)} \left[ AJ\right] _i ,\label{AJ1}
\end{equation}
the condition (\ref{condition}) is {\em not} equivalent to $[AJ]_1=0$.
Here, the criterion becomes that $[AJ]_1$ may not contain any primary
fields. Indeed, if $A$ is a primary with respect to
$T$ (and $J$ a primary of dimension $1$), we see from eq.
(\ref{projection}) with $B=T$ that $[JA]_1$
is primary.  Because the primary at $[JA]_1$ is the same as the one in
$[AJ]_1$, eq. (\ref{condition}) translates in the requirement that there is
no primary in $[AJ]_1$.

\section{Induced and effective $\Ww$ gravities }

Suppose we have a  $\Ww$ algebra with generators $\Phi^i$, then the induced
action $\Gamma$ of the $\Ww$ gravity is defined by
\begin{equation}
Z[{\bf h}] = e^{-\Gamma  \left[ {\bf h}\right] } = \left< e^{
-\frac{1}{\pi}\int {\bf h \cdot \Phi}}\right>.
\label{indact}
\end{equation}
See \cite{STT1} for an extensive account of induced and effective $\Ww$
gravity theories.

Suppose the  $\Ww$ algebra contains a free field $F$ that can be
factored out. We will denote by $\tilde{\Phi }^i$ the
generators (anti-) commuting with $F$. It can easily be shown that one can
invert the algorithms of the previous section. More specifically, we can
write
\begin{equation}
\Phi ^i = \tilde{\Phi}^i + P^i[{\bf \tilde{\Phi}},F]
\end{equation}
where the $P^i[{\bf \tilde{\Phi}},F]$ are some differential polynomials
with all terms at least of order $1$ in $F$. For the fields ${\bf \tilde
\Phi}$ we then define the induced action $\tilde{\Gamma}$ and $\tilde{Z}$
as in (\ref{indact}). The main result of \cite{STT2}, in the case of the
$N=3,4$ superconformal algebras, was that the induced action
$\tilde{\Gamma}\left[ {\bf h}\right] $ of the non-linear supergravity
could be obtained from the linear one by integrating out the free
fields:
\begin{equation}
\tilde{Z}\left[ {\bf h} \right] =\int [d h_{F}]\;Z\left[ {\bf
h_\Phi },h_F\right] . \label{integrateF}
\end{equation}
We will now give a heuristic argument that this should be the case for a
generic $\Ww$ algebra.

We can compute $Z$ as follows
\begin{equation}
Z[{\bf h},h_F] = \left< \exp\left[ -\frac{1}{\pi}\int h_k(\tilde{\Phi}^k +
P^k[{\bf \tilde{\Phi}},F]) + h_F F\right] \right>_{OPE}.
\label{Z}
\end{equation}
We assume that there exists a path integral formulation for this
expression. This could change the form of the polynomials $P^k$, due to
normal ordering problems. Now we integrate (\ref{Z}) over $h_F$, and change
the order of integration. The last term in the exponential gives us $\delta
(F)$, so all terms containing $F$ can be dropped.\footnote{Here we use the
fact that the $\tilde{\Phi}^i$ commute with $F$.} The remaining expression
is exactly $\tilde{Z}$.

Going to the effective theory, we define
\begin{equation}
e^{-W[{\bf t}]} = \int [d{\bf h}]\,
Z[{\bf h}]\exp\left[ {1\over\pi} \int {\bf h \cdot t}\right]
. \end{equation}
{}From relation (\ref{integrateF}), we get
\begin{equation}
\tilde{W}[{\bf t_\Phi}] = W[{\bf t_\Phi}, t_F = 0].\label{defineW}
\end{equation}

Finally,
if a free field $F$ can be integrated out this should reflect itself in
the Ward identities
\begin{equation}
\bar{\partial }\frac{\delta Z}{\delta h_i(x,\bar{x})}=\frac{1}{\pi
}\sum_{j,n}^{}\frac{(-1)^{n-1}}{(n-1)!}\partial ^{n-1}\left(
h_j\left<\left[\Phi^i\Phi ^j\right] _n
e^{-\frac{1}{\pi }\int {\bf h \cdot\Phi}}\right>\right).
\label{wardid}
\end{equation}
Indeed, when we factor out a fermion, the Ward identity corresponding to
$h_i = h_{\psi }$ in (\ref{wardid}) is
\begin{equation}
\bar{\partial} u^\psi = -{\lambda\over\pi} h_{\psi } + F\left[
{\bf h},{\bf u}, u^\psi\right] ,
\end{equation}
where
\begin{equation}
u^i\equiv \frac{\delta \Gamma }{\delta h_i}.
\end{equation}
Setting $u^{\psi }=0$, we can now fill in the solution for $h_{\psi}$ in
the other Ward identities. In this way, the fermion
$\psi $ completely disappears from our theory. It is pretty obvious that
the same can be done for a couple $(\xi ^+,\xi ^-)$ of symplectic bosons,
by looking at the equations with $h_i = \xi ^{\pm}$.

Now suppose we want to solve from these equations a particular source
$h_J$ corresponding to a $U(1)$ field $J$.
The Ward identity (\ref{wardid}) of $h_J$ has an anomalous term
proportional to $\partial h_J$. This means that we will only be able to
remove $J$ if $h_J$ never appears underived.
So our criterion for factoring out of a $U(1)$ field $J$ should be that
in all Ward identities of the theory, the coefficient of $h_J$ vanishes.
If we look at (\ref{wardid}) for some source $h_i$, this term is given by
\begin{equation}
-\frac{1}{\pi }\sum_{n}^{}\frac{(-1)^{n-1}}{(n-1)!}h_J\partial
^{n-1}\left<\left[ \Phi _iJ\right] _n (x) e^{-\frac{1}{\pi
}\int {\bf h \cdot\Phi}}\right>.
\end{equation}
Now we can use (\ref{AJ1}) to simplify this to
\begin{equation}
-\frac{1}{\pi }h_J\left<\left[ J\Phi _i\right] _1 (x)
e^{-\frac{1}{\pi }\int {\bf h \cdot\Phi}}\right>.
\end{equation}
So we see that requiring this term to vanish yields exactly the condition
(\ref{condition})!

\section{Conclusion}
We have given explicit algorithms for factoring out free fields,
including a simple criterion for the factorisation of $U(1)$ fields.  These
algorithms are ideally suited for computer implementation, e.g using the
{\em Mathematica}${}^{TM}$ package OPEdefs \cite{kris}.  We have also shown
that this decoupling is equivalent to integrating out fields from an
induced $\Ww$ gravity theory.  We have worked purely at the quantum
mechanical level, but it is to be expected that analogous algorithms will
exist in the classical case.  In fact, recently \cite{FORT} a number of
classical $\Ww$ algebras were constructed by hamiltonian reduction,
containing bosons of dimensions $1$ and $1/2$ that could be decoupled.  In
\cite{rag} classical $\Ww$ algebras obtained by supersymmetric hamiltonian
reduction \cite{FRS} (based on $OSp(1|2)$ embeddings in affine Lie
superalgebras) were shown to be equivalent to non-supersymmetric ones
(based on $Sl(2)$ embeddings), after factorization of bosonic and fermionic
dimension $1/2$ fields.

\vspace{1cm}
{\bf Acknowledgements}\\
We would like to thank J. M. Figueroa-O'Farrill and W. Troost for helpful
suggestions and comments.


\begin{thebibliography}{MMMM}
\bibitem{GS} P. Goddard and A. Schwimmer, Phys. Lett. {\bf B214} (1988)
209.
\bibitem{FORTW}L. Feh\'er, L. O'Raifeartaigh, P. Ruelle, I. Tsutsui and
A. Wipf, Phys. Rep. {\bf 222} (1992) 1.
\bibitem{BO}M. Bershadsky and H. Ooguri, Comm. Math. Phys.{\bf 126} (1989)
49.
\bibitem{TdB} J. de Boer and T. Tjin, Preprint THU-93-05.
\bibitem{ST} A. Sevrin and W. Troost, Preprint LBL-34125, UCB-PTH-93/19,
KUL-TF-93/21.
\bibitem{FRS} L. Frappat, E. Ragoucy and P. Sorba, Preprint
ENSLAPP-AL-391/92, to appear in Comm. Math. Phys..
\bibitem{bais} F.A. Bais, P. Bouwknegt, K. Schoutens and M. Surridge, Nucl.
Phys. {\bf B304} (1988) 348.
\bibitem{STT2}A. Sevrin, K. Thielemans and W. Troost, Preprint LBL-33778,
UCB-PTH-93/07, KUL-TF-93/10, to appear in Phys. Rev. {\bf D}.
\bibitem{STT1}A. Sevrin, K. Thielemans and W. Troost, Preprint LBL-33738,
UCB-PTH-93/06, KUL-TF-93/09.
\bibitem{kris} K. Thielemans, Int. J. Mod. Phys. {\bf C2} (1991) 787.
\bibitem{FORT}L. Feh\'er, L. O'Raifeartaigh, P. Ruelle and I. Tsutsui,
Preprint BONN-HE-93-14, DIAS-STP-93-02.
\bibitem{rag}E. Ragoucy, Preprint NORDITA 93/39 P.

\end{thebibliography}
\end{document}